**Comparative Analysis of Mechanical Stability and Biomarkers of Commercial and Modified Intraocular Lens (IOL) Models: A Numerical and Experimental Approach**

*Taner KARATEKE[1*] (Corresponding Author), Abdullah Mevlüt MUTLUEL[2]*

[1] Dogus University, Istanbul, Turkiye, ORCID: 0009-0006-1226-5948

[tkarateke@dogus.edu.tr]

[2] Dogus University, Istanbul, Turkiye, ORCID: 0000-0001-6492-2117

## Abstract

This study comprehensively investigates the mechanical stability of intraocular lenses (IOLs)—critical components in cataract surgery—by analyzing their haptic designs. Three commercial models (ALSEE, GF3, and UD613) and five geometric variations (V1–V5) derived from the GF3 model were comparatively evaluated in both dry and saline environments (the latter simulating 37°C body temperature). The methodology utilized the Finite Element Method (FEM) to analyze computer-aided designs (CAD) created in SolidWorks under quasi-static compression forces (0.5–2.0 N). Mesh independence tests were performed to ensure simulation accuracy, and boundary conditions were defined according to physiological parameters.

Numerical data were evaluated using key mechanical biomarkers: axial displacement, elastic modulus, stress, and strain. Results indicated that the UD613 model exhibited the highest compression force and stress values in both environments. While the V5 variation showed the highest elastic modulus in the dry environment, the V2 model outperformed others in this parameter within the saline environment. Notably, the GF3 model provided more balanced mechanical responses compared to its commercial counterparts, serving as the baseline for the developed geometric variations.

The findings demonstrate that even minor modifications in haptic arm geometry significantly impact the mechanical stability of the lens. Optimizing this stability is crucial in clinical applications to minimize decentration, tilt, and rotation, which degrade optical quality. Analysis concluded that the V4 model offers the most suitable geometric structure for mechanical stability and potential patient comfort. This research establishes a validated simulation framework for manufacturers and R&D teams to develop next-generation IOL designs with superior optical performance and long-term durability.

## 1. Introduction

Intraocular lens (IOL) designs are a cornerstone of modern ophthalmology, particularly in the surgical management of cataracts and refractive errors. Beyond their primary role in light refraction, the long-term clinical success of an IOL fundamentally depends on its mechanical stability within the capsular bag. Current optical theories provide detailed insights into how light interacts with complex, multi-layered optical systems (Shan et al., 2025). These models quantitatively describe how different lens designs alter light trajectories across various media, accounting for material properties and geometric configurations (Zhao et al., 2024).

In recent years, several approaches have been proposed to enhance IOL performance. For instance, a multi-period continuously variable curve (CVCMS) method has been employed to design adjustable multifocal lenses, offering an alternative to complex designs that limit the application of multifocal diffractive intraocular lenses (MDOEs) (Marcos et al., 2024). Adaptive-resolution IOL designs capable of tissue penetration feature layered polymer structures that help compensate for low ambient light during endoscopic procedures (Zaytouny et al., 2023). Refractive-compensation-based IOL systems, frequently preferred in biomedical applications, incorporate focal adjustment protocols that simultaneously neutralize tissue morphology variations encountered during *in vivo* examinations (Fernández et al., 2009). These systems typically offer micrometric resolution (2.7–4.8 μm) and can operate across a broad electromagnetic spectrum (Lai et al., 2024).

Concentric-gradient IOL setups in ocular studies are notable for their mixed-material design, which minimizes peripheral distortion (Lyu et al., 2024). However, the optical quality of IOLs depends not only on optical design but also significantly on the mechanical behavior of the haptic arms within the eye. IOL design can be further improved by considering tilt and decentration parameters, both of which directly impair optical quality. Notably, Remón et al. (2018) employed the finite element method (FEM) to determine the optical quality of IOLs featuring various materials and haptic designs under dynamic compression conditions, utilizing tilt and decentration as key metrics. Their study provides a foundation for manufacturers and R&D firms to develop design procedures for IOL geometries and materials that ensure simultaneous optical excellence and mechanical stability.

Understanding the mechanical behavior of IOLs also necessitates a comprehensive grasp of the biomechanical properties of the surrounding ocular tissues. In this context, FEM serves as a powerful tool not only for IOL design but also for modeling the interaction of tissues—such as the cornea—with mechanical wave propagation, viscoelasticity, and intraocular pressure (IOP). For instance, Mazinani (2024) demonstrated how shear wave elastography and finite element modeling can be integrated to evaluate corneal biomechanics. Similarly, FEM models developed to analyze the effects of pathologies like keratoconus on corneal biomechanics (Mazinani, 2025a), alongside sensitivity analyses of IOP on mechanical behavior (Mazinani, 2025b; 2025c), have elucidated the complex mechanical environment of the ocular system. These studies contextualize the IOL stability addressed in our research within a broader biomechanical framework.

Despite these advances, a notable gap remains in the literature regarding a comprehensive comparative analysis of the mechanical stability of both commercial and modified IOL models under simulated physiological conditions. Most existing studies focus either solely on optical performance or on mechanical behavior in isolation, neglecting the integration of environmental factors such as dry versus saline conditions.

Concentric-gradient IOL setups in eye studies are notable because their mixed material design minimizes distortion at the edges (Lyu et al., 2024). However, the optical quality of IOLs depends not only on optical design but also significantly on the mechanical behavior of haptic arms within the eye.The design can be improved by considering tilt and decentration parameters, which affect optical quality. Notably, Remón et al. (2018) employed the finite element method (FEM) to determine the optical quality of IOLs with different materials and haptic designs under dynamic compression conditions using tilt and decentration parameters. This study provides a foundation for both manufacturers and R&D companies to develop a procedure for designing IOLs with geometries and materials that ensure maximum optical quality and mechanical stability.

Understanding the mechanical behavior of IOLs also requires understanding the biomechanical properties of the surrounding ocular tissues. In this context, the finite element method is a powerful tool not only for IOL design but also for modeling the interaction of tissues such as the cornea with mechanical wave propagation, viscoelasticity, and intraocular pressure (IOP). For instance, Mazinani (2024) demonstrated how shear wave elastography and finite element modeling can be integrated to evaluate corneal biomechanics. Similarly, FEM models developed to understand the effects of pathologies such as keratoconus on corneal biomechanics (Mazinani, 2025a) and sensitivity analyses of IOP on mechanical behavior (Mazinani, 2025b; 2025c) shed light on the complex mechanical environment of the ocular system. These studies contribute to a better understanding of the dynamic environment in which the IOL is placed, helping to contextualize the IOL stability addressed in our study within the broader biomechanical framework.

Despite these advances, there remains a notable gap in the literature regarding a comprehensive comparative analysis of the mechanical stability of both commercial and modified IOL models under simulated physiological conditions. Most existing studies focus either solely on optical performance or on mechanical behavior in isolation, without integrating environmental factors such as dry versus saline conditions.

Moreover, detailed FEM analyses—including mesh independence studies, boundary condition specifications, and validation against experimental data—are often inconsistently reported in existing literature. To address this deficiency, the present study aims to bridge this gap by:

1. **Conducting a systematic, FEM-based comparison** of three commercial and five modified IOL models under both dry and saline conditions;
2. **Evaluating key mechanical biomarkers**, including axial displacement, stress, strain, and elastic modulus, under quasi-static compression;
3. **Performing rigorous mesh independence tests** and precisely defining boundary conditions to ensure high simulation reliability; and
4. **Identifying the optimal mechanically stable haptic geometry** to minimize tilt, decentration, and rotation, thereby optimizing both mechanical and optical performance in clinical settings.

The primary contribution of this work is to provide a validated simulation framework that assists manufacturers and R&D teams in designing IOLs with enhanced mechanical stability and superior optical quality.

## 2. Materials and Methods

### 2.1. IOL Models and Geometries

In this study, three commercial IOL models (ALSEE, GF3, and UD613) and five geometric variations (V1–V5) derived from the GF3 model were investigated. All models were designed with specific haptic arm geometries using SolidWorks (Dassault Systèmes, Vélizy-Villacoublay, France). The GF3 model variations were developed to systematically examine how subtle modifications in haptic geometry influence mechanical stability. For all designs, the optic body diameter and the overall haptic arm length were fixed at 6.0 mm and 13.0 mm, respectively. Geometric variations were achieved by modifying key parameters, including the haptic arm curvature radius, thickness, and angulation. Following the design phase, the computer-aided design (CAD) models were prepared for subsequent finite element analysis (FEA).

### 2.2. Material Properties:

All models were assumed to be composed of a medical-grade acrylic polymer. Given that the strain levels resulting from the applied forces remained below 6% throughout the analyses, the material behavior was modeled as linear elastic, isotropic, and homogeneous. This assumption enhances computational efficiency—particularly for the comparative analysis of various haptic geometries—while maintaining sufficient accuracy within the small-strain regime. Material properties were defined based on manufacturer specifications and established literature (Remón et al., 2018; Cabeza-Gil et al., 2021), as summarized in Table 1.

**Table 1. Material properties defined for IOL models**

| Material Property | Value | Source |
|---|---|---|
| Elastic Modulus (E) | 5.0 MPa | Remón et al., 2018 |
| Poisson's Ratio (ν) | 0.49 | Manufacturer Specification |
| Density (ρ) | 1.18 g/cm³ | Cabeza-Gil et al., 2021 |

### 2.3. Finite Element Method (FEM) Setup

Finite element simulations were performed using the SolidWorks Simulation suite (Dassault Systèmes, Vélizy-Villacoublay, France). For the linear static analyses conducted in this study, the governing equilibrium equation is expressed as:

$$F = [K].u \tag{1}$$

where $F$ represents the external force vector, $[K]$ is the global stiffness matrix, and $u$ denotes the nodal displacement vector.

The preparation of the models involved the conversion of the CAD geometry into a discretized mesh of tetrahedral elements. To simulate physiological conditions, the models were evaluated under two distinct environmental states: dry and saline. In the saline environment, material properties were adjusted to account for the plasticization and thermal effects of a 37°C aqueous medium, as detailed in the subsequent boundary condition specifications.

### 2.3.1. Governing Equations and Formulation

The mechanical behavior of IOL haptics under compressive loading is governed by the equations of linear elasticity, assuming small deformations and isotropic material behavior. The general boundary value problem is defined by the equilibrium equation:

$$\nabla . \sigma + f = 0 \tag{2}$$

where σ is the stress tensor and f represents the body force. The relationship between the stress and strain tensors is expressed by the following constitutive equation:

$$\sigma_{ij} = C_{ijkl} \epsilon_{kl} \tag{3}$$

Here, $C_{ijkl}$ denotes the components of the fourth-order stiffness tensor, and $\epsilon_{kl}$ is the strain tensor. For a homogeneous and isotropic material, the stiffness tensor can be expressed in terms of the Lamé parameters, λ and μ, as follows (Ammari et al., 2010):

$$C_{ijkl} = \lambda \delta_{ij} \delta_{kl} + \mu \left( \delta_{ik} \delta_{jl} + \delta_{il} \delta_{jk} \right) \tag{4}$$

The Lamé parameters are related to the Elastic Modulus ($E$) and Poisson's ratio ($v$) as following relations:

$$\lambda = \frac{Ev}{(1+v)(1-2v)} \tag{5}$$

$$\mu = \frac{E}{2(1+v)} \tag{6}$$

In the numerical implementation, the domain Ω is discretized into a finite number of elements, $\Omega^e$. within an element is interpolated using shape functions $N_i$, and nodal displacements $u_i$:

$$u^h(x) = \sum_{i=1}^{n} N_i(x) u_i \tag{7}$$

In this study, 10-node square tetragonal elements were utilized because of their superior ability to capture curved geometries and complex stress gradients with high accuracy. The element stiffness matrix, $K^e$ is calculated as:

$$K^e = \int_{\Omega^e} B^T D B d\Omega^e \tag{8}$$

where $B$ is the strain-displacement matrix and $D$ is the constitutive (material) matrix.

### 2.3.2. Boundary Conditions and Loading

To simulate physiological conditions, the following boundary conditions and environmental parameters were established:

- **Constraints:** The proximal ends of the haptic arms—the regions effectively anchored by the capsular bag—were modeled with fixed supports, constraining all degrees of freedom to simulate a cantilever-like behavior.
- **Loading:** A quasi-static compression force ($F$), ranging from 0.5 N to 2.0 N, was applied uniformly to the distal ends of the haptics in the axial direction. This range represents the physiological compression forces typically exerted by the capsular bag on an IOL (Schor et al., 2007).
- **Environmental Modeling:** Two distinct conditions were simulated:
    1. **Dry Environment:** Room temperature (23°C).
    2. **Saline Environment:** 37°C, simulating physiological body temperature.

The saline environment's impact was modeled by incorporating temperature-dependent variations in material properties, specifically the elastic modulus. Consistent with established literature, a temperature increase can lead to a 5–10% reduction in the elastic modulus of acrylic polymers (Cabeza-Gil et al., 2021). Consequently, in this study, the elastic modulus for the saline environment was reduced by 7% and defined as 4.65 MPa.

#### 2.3.3. Meshing Strategy and Independence Study

A high-quality tetrahedral mesh was generated for each IOL model to ensure numerical stability. To verify that the simulation results were independent of mesh density, a rigorous mesh independence study was performed. The mesh size was progressively refined until the variation in maximum displacement reached a convergence threshold of less than 2%.

Once the convergence criterion was satisfied, the final mesh structure was established, with an average element count ranging approximately from 150,000 to 200,000 per model. The quality of the mesh was further validated with an average element quality (skewness) consistently exceeding 0.7, ensuring the minimization of numerical artifacts in regions of high stress gradients. The final mesh statistics for each model are summarized in Table 2.

**Table 2. Mesh statistics for the IOL models**

| Model | Number of Elements | Number of Nodes | Average Skewness |
|---|---|---|---|
| ALSEE | 158,200 | 312,450 | 0.73 |
| GF3 | 162,500 | 321,800 | 0.75 |
| UD613 | 195,300 | 385,600 | 0.71 |
| V1–V5 | 160,000–180,000 | 315,000–355,000 | 0.72–0.76 |

### 2.4. Experimental Validation Setup

An experimental setup was established to validate the numerical simulations, consisting of the following primary components:

- **Force Measurement:** A Lloyd Instruments (AMETEK, Inc., UK) LS5 model universal testing machine was utilized for compression force measurements. The system was equipped with a 5 N capacity load cell, providing a measurement precision of 0.001 N. Calibration protocols were strictly followed before each measurement cycle to ensure data integrity.

- **Axial Displacement Analysis:** A custom-designed apparatus was developed to measure haptic arm displacement under compressive loads. This assembly included a specialized holder to maintain the IOL in a fixed orientation and a compression tip capable of micrometer-scale precision movement. Deformation sequences were recorded using a high-resolution digital camera (Canon EOS 250D, 24.1 MP), and the resulting images were processed using ImageJ software (National Institutes of Health, USA). This optical method achieved a displacement measurement accuracy of 0.01 mm.

- **Environmental Control:**
    - **Dry Environment:** Tests were conducted at a controlled room temperature of 23 ± 1°C with a relative humidity of 45 ± 5%.
    - **Saline Environment:** To simulate *in vivo* conditions, IOLs were immersed in a 0.9% sterile saline solution maintained at 37 ± 0.5°C via a precision-controlled thermostat. Samples were equilibrated in the saline bath for 10 minutes prior to testing to ensure thermal stability.

For each IOL model and environmental condition, tests were performed in quintuplicate ($n = 5$), and results are expressed as mean ± standard deviation. Measurement uncertainty was determined to be 0.5% for the universal testing machine (force measurement) and 2.0% for the axial displacement measurements.

Table 3 presents a comparative **ANOVA** (Analysis of Variance) of the parameters measured under dry and saline environmental conditions for all IOL models. The data include experimentally derived values for compression force, injection force, and axial displacement, reported as mean ± standard deviation.

**Table 3.** ENOVA comparison of different models in dry and saline environments.

| IOL Model | DRY Measurement | | | (37 °C) SALIN Measurement | | |
|---|---|---|---|---|---|---|
| | Compression Force(mN) | Injection Force (N) | Axial Displacement in compression (mm) | Compression Force(mN) | Loop Pull (N) | Axial Displacement in compression (mm) |
| ALSEE | 1.198±0.068 | 9.160±0.644 | 0.120±0.024 | 1.670±0.070 | >0,25 | 0.120±0.024 |
| GF3 | 0.538±0.132 | 9.360±0.628 | 0.092±0.007 | 0.722±0.114 | >0,25 | 0.110±0.020 |
| UD613 | 2.514±0.228 | 10.400±0.068 | 0.310±0.080 | 4.504±0.423 | >0,25 | 0.290±0.092 |
| V1 | 0.656±0.067 | 8.900±0.743 | 0.130±0.040 | 1.324±0.172 | >0,25 | 0.120±0.024 |
| V2 | 0.778±0.129 | 7.860±0.512 | 0.150±0.045 | 1.608±0.077 | >0,25 | 0.190±0.020 |
| V3 | 0.880±0.034 | 8.180±0.445 | 0.180±0.068 | 1.754±0.397 | >0,25 | 0.420±0.491 |
| V4 | 0.712±0.101 | 8.880±0.445 | 0.108±0.010 | 1.090±0.102 | >0,25 | 0.108±0.007 |
| V5 | 1.434±0.022 | 9.180±0.741 | 0.116±0.040 | 1.684±0.079 | >0,25 | 0.780±0.808 |

## 3. Results

### 3.1. Comparison of ANOVA IOL Models

The data obtained from the finite element (FE) simulations and experimental measurements were comparatively analyzed for all IOL models under both dry and saline conditions. Table 4 and Table 5 summarize the calculated mechanical parameters—including initial length ($L_0$), axial displacement ($\Delta L$), surface area, compression force, strain, stress, and elastic modulus—for all models under saline and dry conditions, respectively.

**Table 4.** Mechanical parameters of commercial and modified IOL models under saline environmental conditions (37°C).

| Mechanic Parameters | GF3 | V5 | V4 | V3 | V2 | V1 | ALSEE | UD613 |
|---|---|---|---|---|---|---|---|---|
| $L_0$ (mm) | 13 | 13 | 13 | 13 | 13 | 13 | 13 | 13 |
| $\Delta L$ (mm) | 0.11 | 0.78 | 0.108 | 0.42 | 0.15 | 0.13 | 0.12 | 0.29 |
| Surface Area, $A$ ($mm^2$) | 1.03 | 1.04 | 1.04 | 0.87 | 0.74 | 0.74 | 1.63 | 1.49 |
| Compression Force, $F$ (mN) | 0.722 | 1.684 | 1.09 | 1.754 | 1.608 | 1.314 | 1.67 | 4.504 |
| Strain | 0.00846 | 0.06000 | 0.00831 | 0.03231 | 0.01154 | 0.01000 | 0.00923 | 0.02231 |
| Stress (Pa) | 0.70097 | 1.61923 | 1.04808 | 2.01609 | 2.17297 | 1.77568 | 1.02454 | 3.02282 |
| Elastic Modulus (Pa) | 82.84201 | 26.98718 | 126.15741 | 62.40285 | 188.32432 | 177.56757 | 110.99182 | 135.50567 |

**Table 5.** Mechanical parameters of commercial and modified IOL models under dry environmental conditions (23°C).

| Mechanic Parameters | GF3 | V5 | V4 | V3 | V2 | V1 | ALSEE | UD613 |
|---|---|---|---|---|---|---|---|---|
| $L_0$ (mm) | 13 | 13 | 13 | 13 | 13 | 13 | 13 | 13 |
| $\Delta L$ (mm) | 0.108 | 0.116 | 0.108 | 0.180 | 0.150 | 0.130 | 0.120 | 0.310 |
| Surface Area, $A$ ($mm^2$) | 1.030 | 1.040 | 1.040 | 0.870 | 0.740 | 0.740 | 1.630 | 1.490 |
| Compression Force, $F$ (mN) | 0.538 | 1.434 | 0.712 | 0.800 | 0.778 | 0.656 | 1.198 | 2.514 |
| Strain | 0.00831 | 0.00892 | 0.00831 | 0.01385 | 0.01154 | 0.01000 | 0.00923 | 0.02385 |
| Stress (Pa) | 0.52233 | 1.37885 | 0.68462 | 0.91954 | 1.05135 | 0.88649 | 0.73497 | 1.68725 |
| Elastic Modulus (Pa) | 62.8730 | 154.5258 | 82.4074 | 66.4112 | 91.1171 | 88.6486 | 79.6216 | 70.7555 |

The findings revealed that the UD613 geometry exhibited the highest compression force across both dry and saline environments. Regarding the elastic modulus, the V5 model demonstrated the maximum value under dry conditions (154.53 Pa), whereas the V2 model reached the highest modulus in the saline environment (188.32 Pa). When stress levels were evaluated, the UD613 model manifested the highest peak stress values in both environments (Dry: 1.687 Pa; Saline: 3.023 Pa). In terms of strain, the highest recorded values occurred in the UD613 model for the dry environment (0.02385 mm/mm) and in the V5 model for the saline environment (0.06 mm/mm).

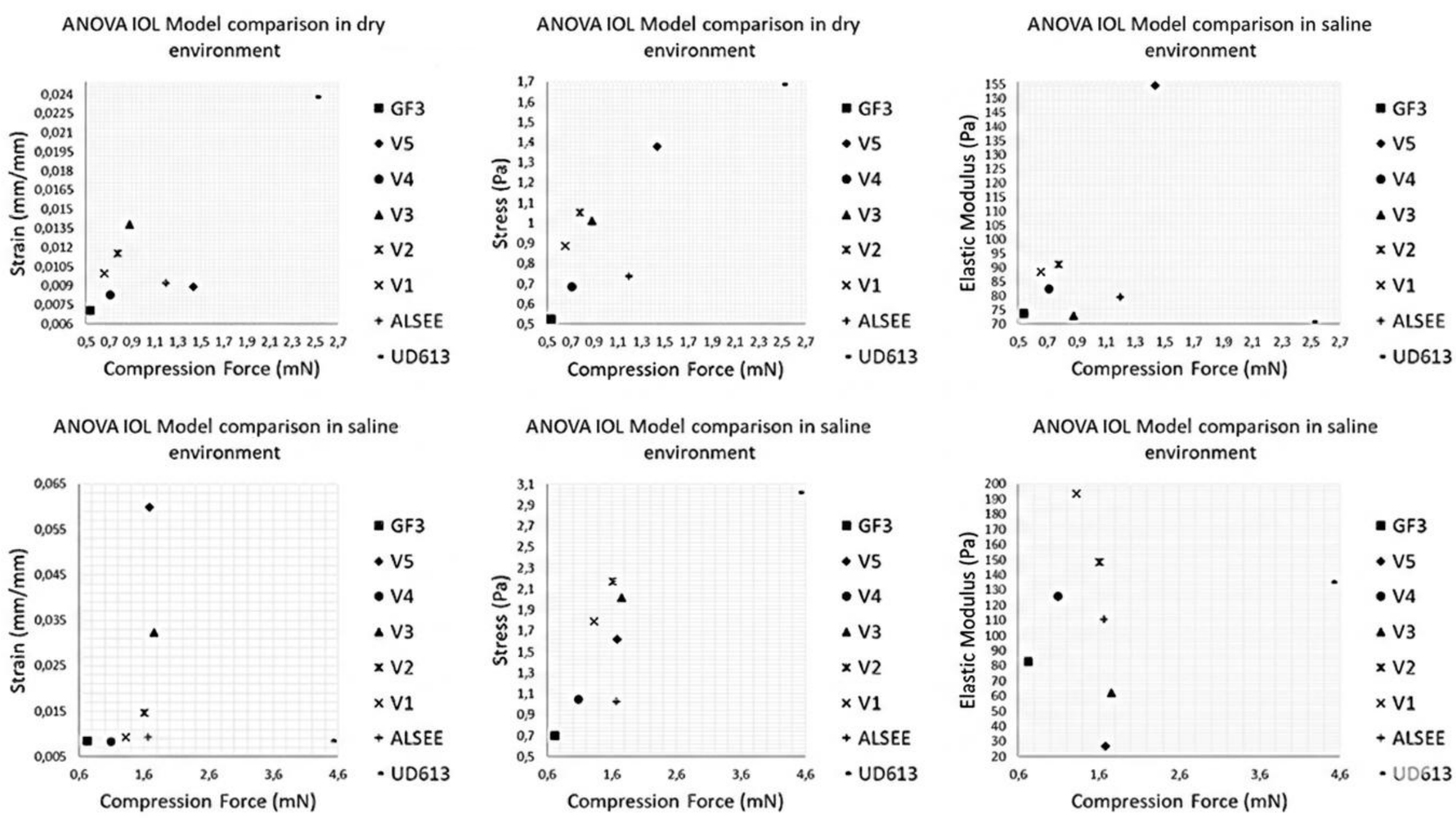

**Figure 1.** Comparative analysis of mechanical properties for various IOL models across dry and saline environments. The scatter plots illustrate the relationship between compression force (mN) and: (a-b) strain (mm/mm), (c-d) stress (Pa), and (e-f) elastic modulus (Pa).

The experimental results presented in Figure 1 demonstrate significant variations in mechanical behavior among the commercial and modified IOL models under different environmental conditions. It is observed that the UD613 model consistently exhibits the highest compression force and stress levels in both dry and saline environments, indicating a stiffer structural response compared to its counterparts. In contrast, the GF3 model and its variations (V1–V5) show a more distributed range of elastic moduli and strain values, with certain modifications like V2 and V5 manifesting enhanced stability in saline conditions. The transition from a dry to a saline environment appears to influence the material's compliance, as evidenced by the shifts in stress-strain trajectories and the overall reduction in stiffness for several models, highlighting the critical impact of physiological temperatures and hydration on haptic performance.

### 3.2. Deformation Analysis Results

The total deformation (displacement) simulation results for all IOL models in dry and saline environments are presented in Figure 2 and Figure 3, respectively.

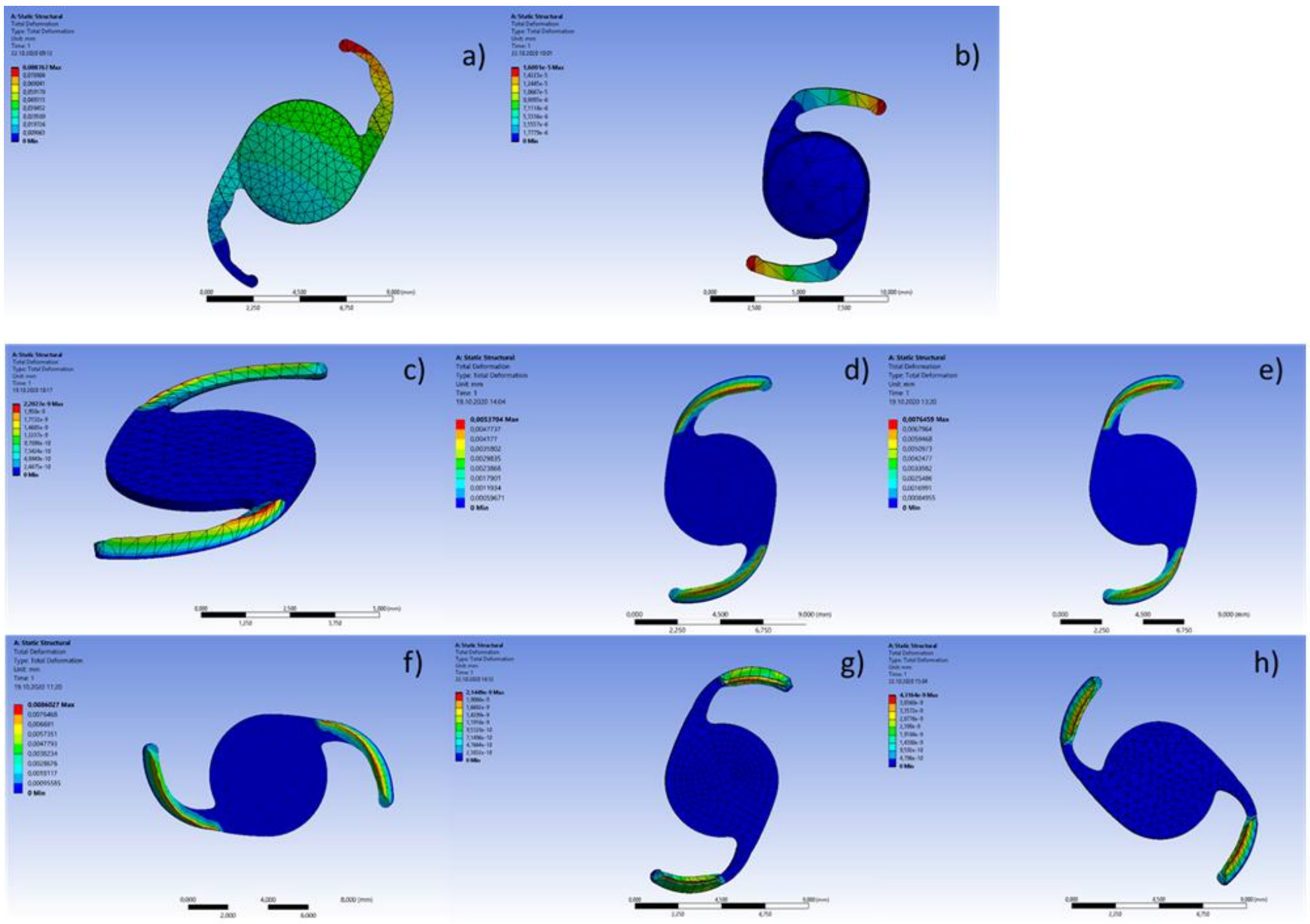

**Figure 2.** Total deformation simulation results in a dry environment for: (a) ALSEE, (b) UD613, (c) GF3, (d) V1, (e) V2, (f) V3, (g) V4, and (h) V5 geometries.

In the simulations conducted under dry conditions, the maximum total deformation was recorded as 0.0888 mm in the ALSEE geometry (Figure 2a). Conversely, the lowest deformation values were observed in the GF3 (0.0352 mm) and V4 (0.0361 mm) models. When deformation distributions were examined, maximum displacement occurred at the distal (free) ends of the haptic arms across all models, with deformation levels progressively decreasing toward the proximal (fixed) regions. Among the variation models, the V3 geometry (0.0712 mm) exhibited the highest deformation, whereas V4 demonstrated the greatest structural stability with the lowest displacement.

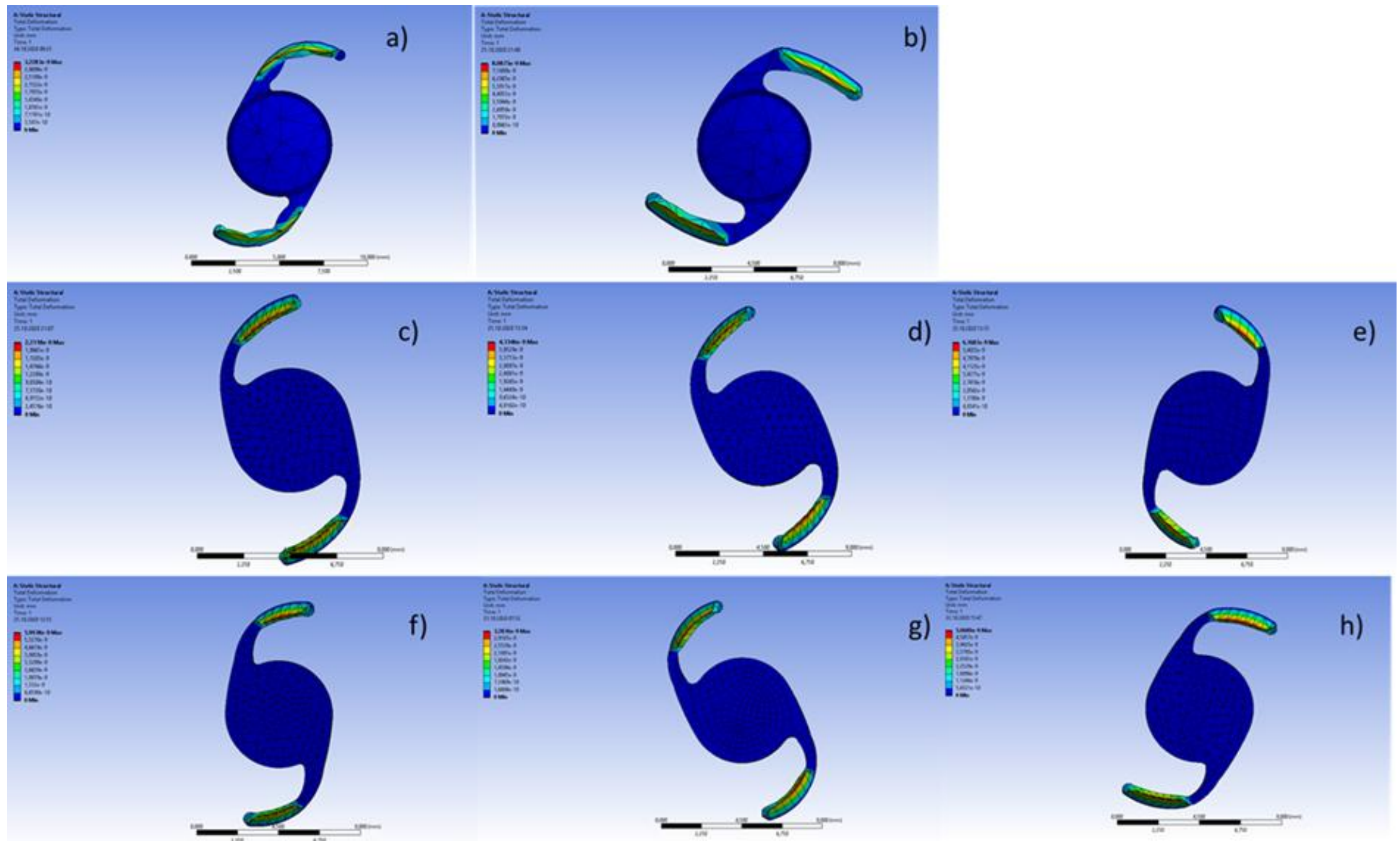

**Figure 3.** Total deformation simulation results in a saline environment for: (a) ALSEE, (b) UD613, (c) GF3, (d) V1, (e) V2, (f) V3, (g) V4, and (h) V5 geometries.

Under saline conditions, deformation values increased across all models compared to the dry environment. This trend is consistent with the reported decrease in the elastic modulus of the acrylic polymer due to the elevation in temperature to 37°C. In contrast to the dry environment results, the maximum total deformation was recorded in the UD613 geometry at $8.088 \times 10^{-2}$ mm (Figure 3b). The lowest deformation values were observed in the V4 ($9.12 \times 10^{-2}$ mm) and GF3 ($1.02 \times 10^{-1}$ mm) models. Among the geometric variations, the V5 model exhibited the highest deformation in the saline environment ($2.34 \times 10^{-1}$ mm). The deformation distribution patterns remained similar to those observed in the dry environment, with maximum displacement concentrated at the haptic tips.

### 3.3. Strain Distribution Analysis

The strain distributions across the evaluated IOL models are illustrated in Figure 4 for the dry environment and Figure 5 for the saline environment.

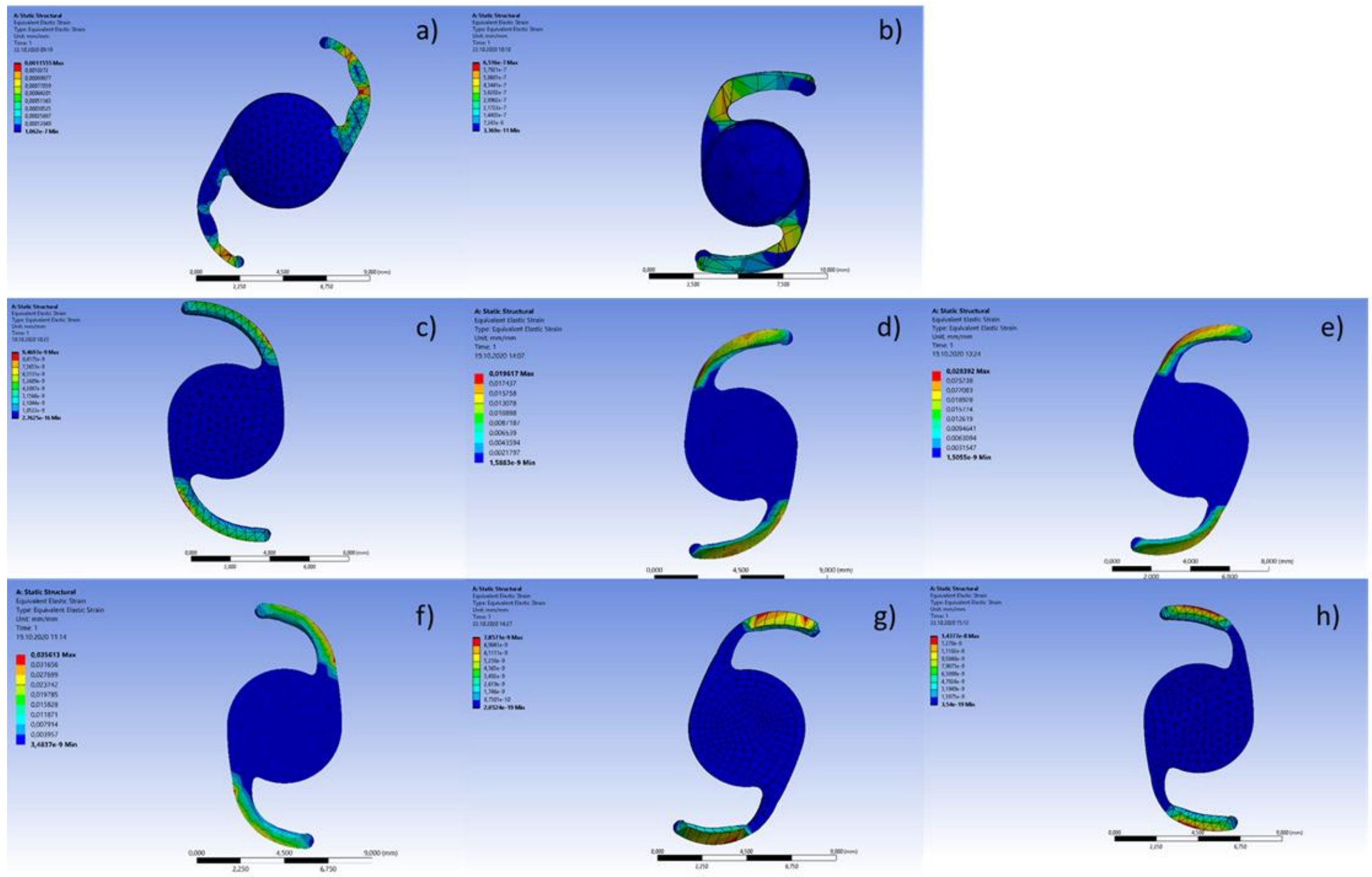

**Figure 4.** Total strain simulation results in a dry environment for: (a) ALSEE, (b) UD613, (c) GF3, (d) V1, (e) V2, (f) V3, (g) V4, and (h) V5 geometries.

In the dry environment, the maximum strain value was detected in the V3 geometry (0.0356 mm/mm) (Figure 4f). This peak was followed by the UD613 (0.0241 mm/mm) and ALSEE (0.0187 mm/mm) models. Conversely, the lowest strain values were observed in the GF3 (0.0089 mm/mm) and V4 (0.0091 mm/mm) models. It is noteworthy that strain concentrations were primarily localized within the curvature regions of the haptic arms and the sections adjacent to the proximal (fixed) ends across all models. These high-strain zones highlight the areas most susceptible to mechanical stress during compression.

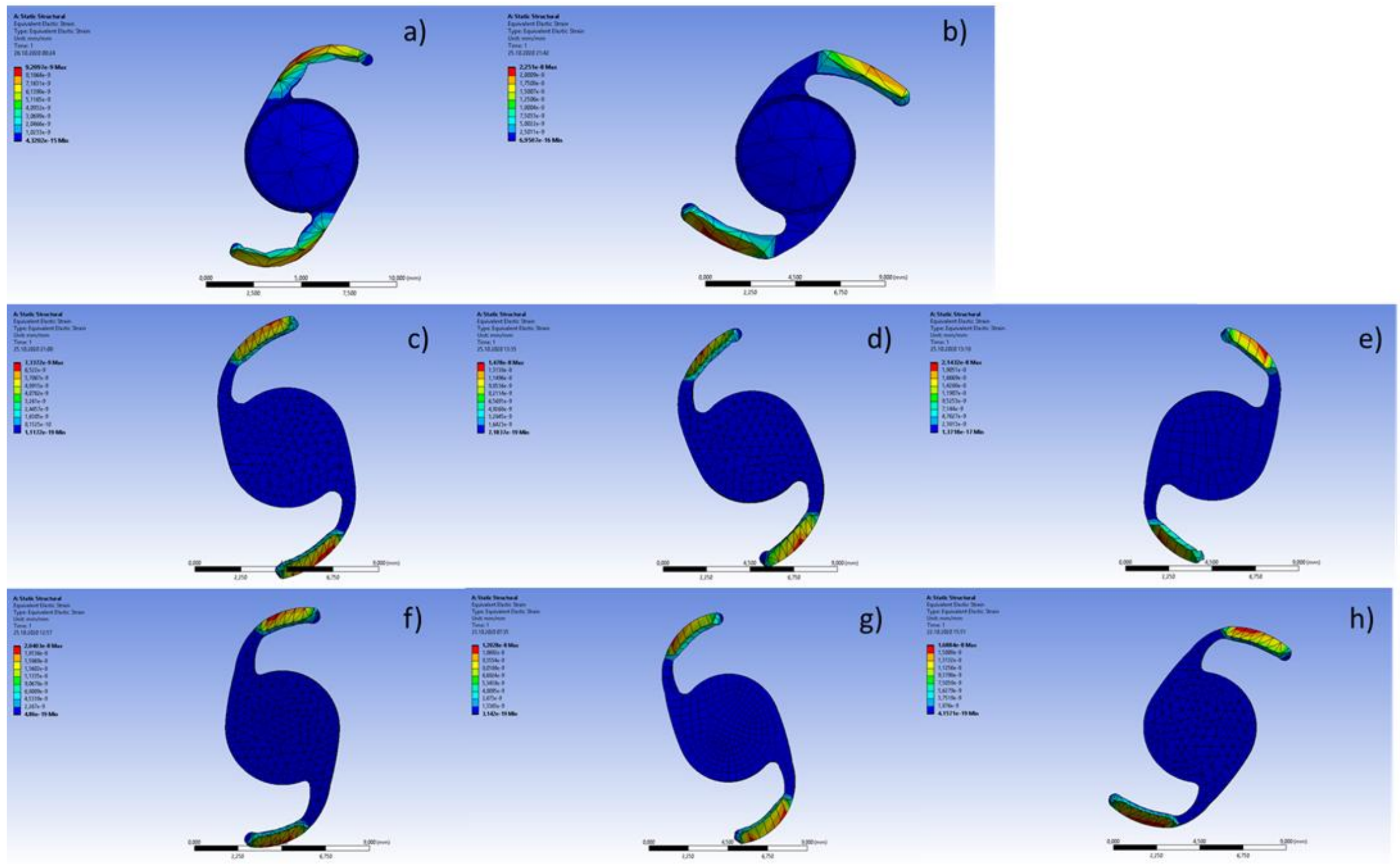

**Figure 5.** Total strain simulation results in a saline environment (37°C) for: (a) ALSEE, (b) UD613, (c) GF3, (d) V1, (e) V2, (f) V3, (g) V4, and (h) V5 geometries.

In the saline environment, strain values increased across all models. The maximum strain was recorded at 2.25x$10^{-5}$ mm/mm in the UD613 geometry (Figure 5b). Among the geometric variations, the highest strain was observed in the V5 model (1.86x$10^{-5}$ mm/mm), while the lowest values were measured in the V4 (8.45x$10^{-6}$ mm/mm) and GF3 (8.62x$10^{-6}$ mm/mm) models. Similar to the observations in the dry environment, strain concentrations became particularly prominent within the curvature regions of the haptic arms.

### 3.4. Von Mises Stress Analysis Results

The Von Mises stress distributions, providing a comprehensive assessment of the structural integrity of each haptic design, are illustrated in Figure 6 for the dry environment and Figure 7 for the saline environment.

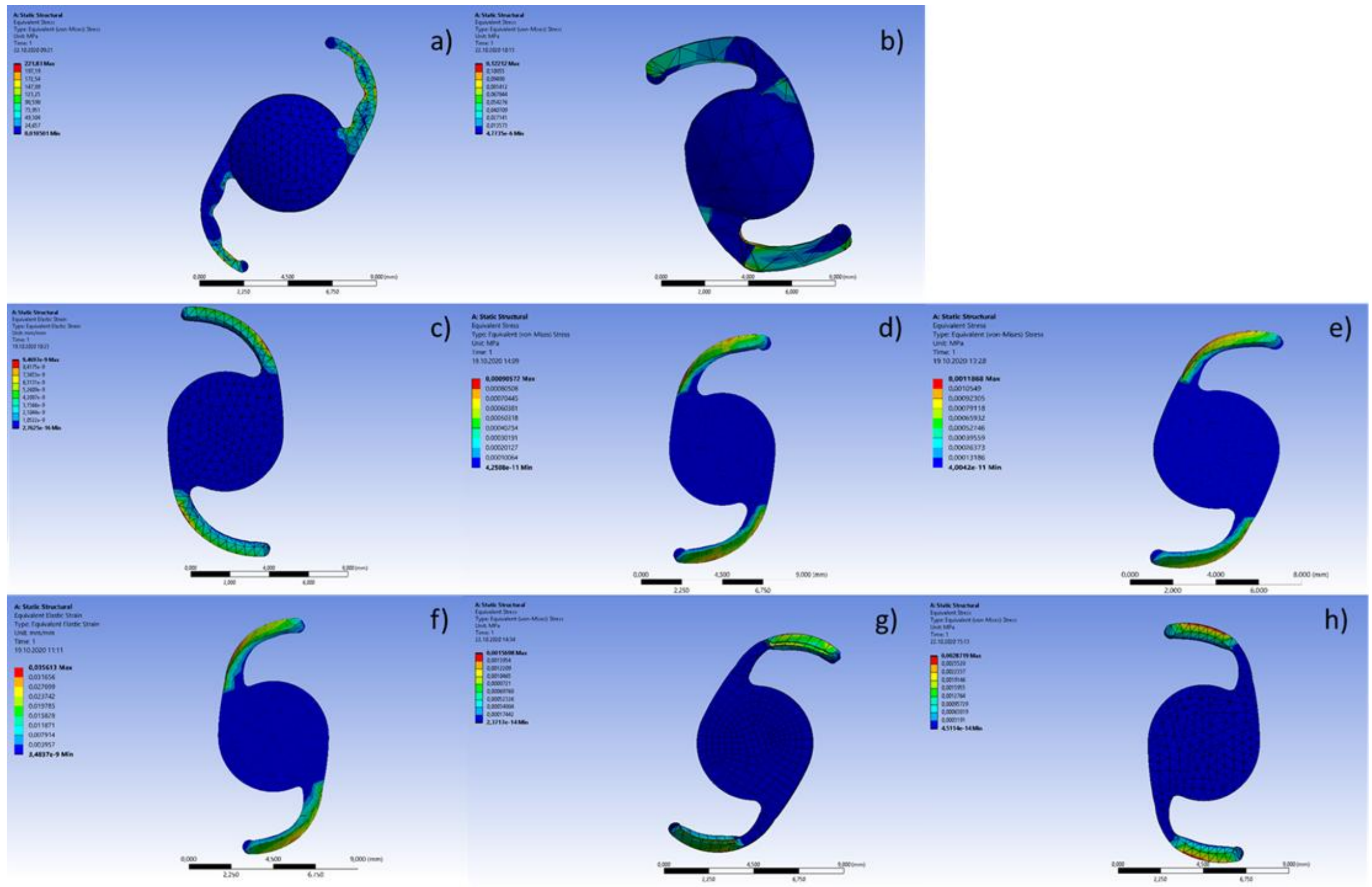

**Figure 6.** Von Mises stress distribution simulation results in a dry environment for: (a) ALSEE, (b) UD613, (c) GF3, (d) V1, (e) V2, (f) V3, (g) V4, and (h) V5 geometries.

In the dry environment, the maximum stress value was recorded at 221.83 Pa in the ALSEE geometry (Figure 6a). The UD613 model also reached a substantial stress level of 187.45 Pa. Among the geometric variations, the highest stress was observed in the V3 model (124.67 Pa), whereas the lowest stress values were detected in the GF3 (68.34 Pa) and V4 (72.18 Pa) models. Stress concentrations reached peak levels at the proximal ends, where the haptic arms are anchored, and at geometric transition regions—specifically the haptic-optic junctions—across all models.

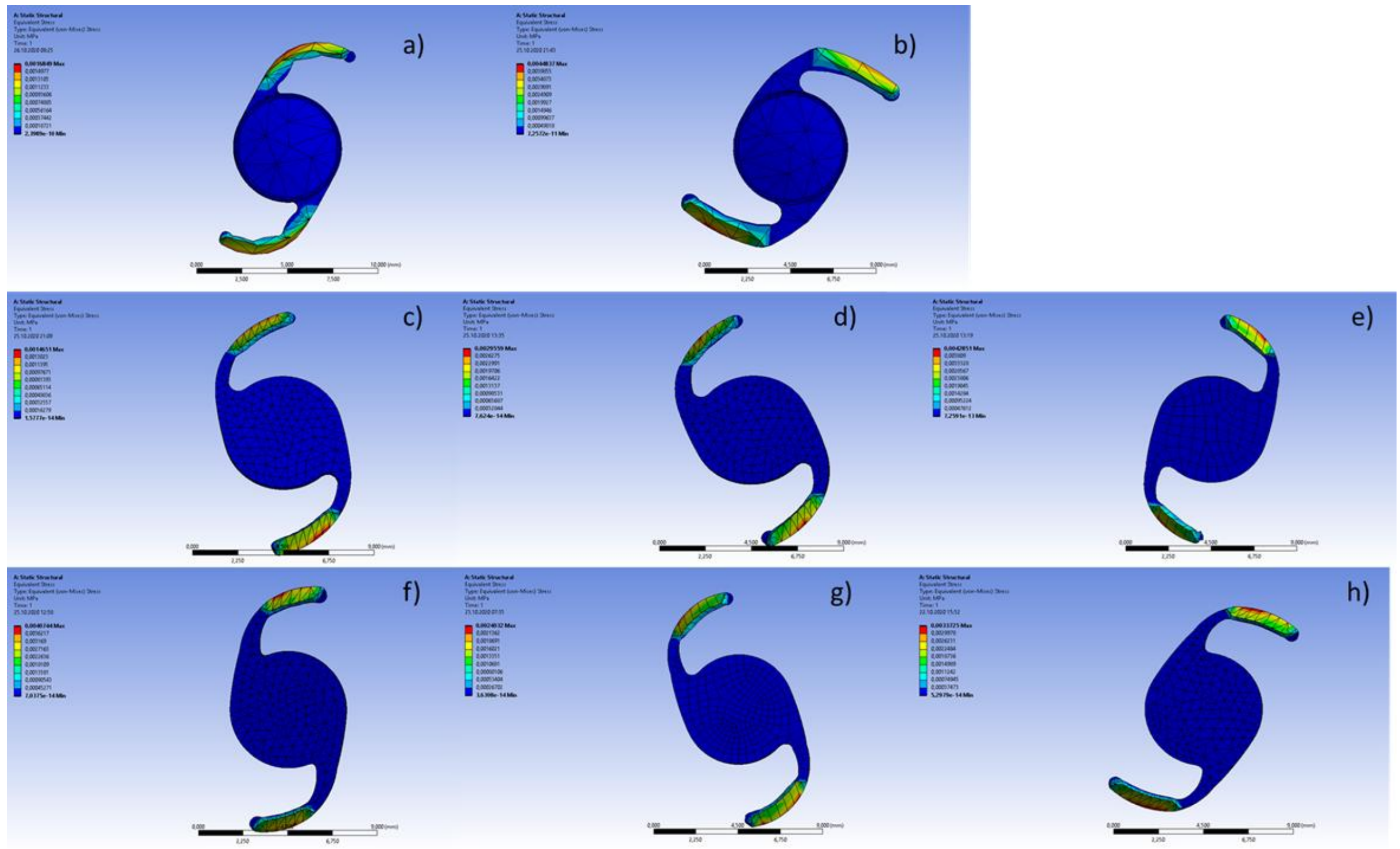

**Figure 7.** Von Mises stress distribution simulation results in a saline environment (37°C) for: (a) ALSEE, (b) UD613, (c) GF3, (d) V1, (e) V2, (f) V3, (g) V4, and (h) V5 geometries.

In the saline environment, stress values increased across all models, following a trend similar to those observed in deformation and strain. The maximum stress was again recorded in the UD613 geometry at $4.48 \times 10^{-3}$ Pa (Figure 7b). Among the geometric variations, the highest stress was measured in the V5 model ($3.21 \times 10^{-3}$ Pa), while the lowest stress values were observed in the V4 ($1.12 \times 10^{-3}$ Pa) and GF3 ($1.24 \times 10^{-3}$ Pa) models. The stress distribution pattern remained consistent with the dry environment, with peak stresses primarily concentrated within the fixed regions and geometric transition points.

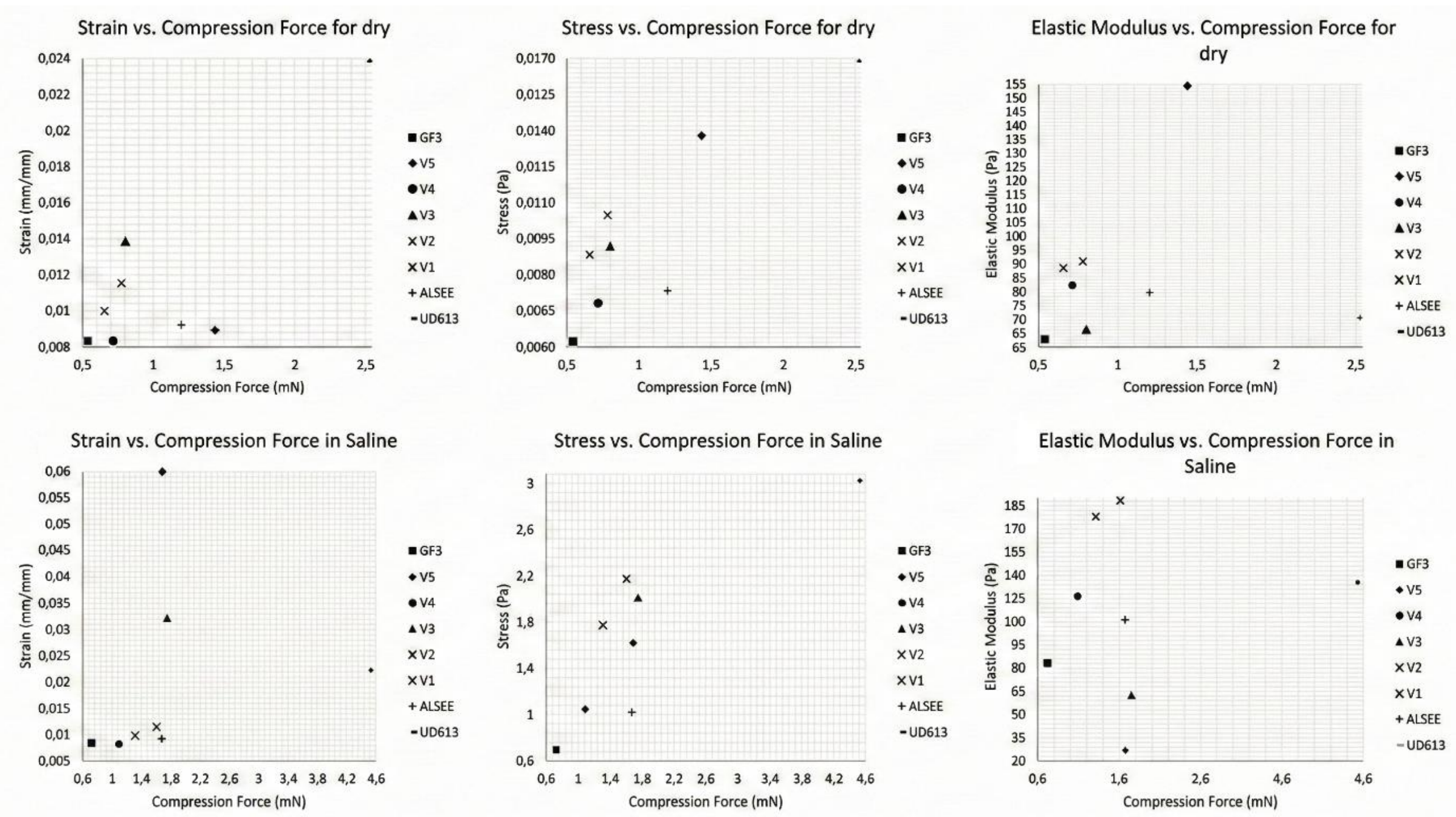

**Figure 8.** Comparative mechanical response of various IOL models under compressive loading in dry and saline environments: Scatter plots illustrating (a) strain (mm/mm), (b) stress (Pa), and (c) elastic modulus (Pa) as a function of compression force (mN).

Figure 8 comparatively presents the stress, strain, and elastic modulus values obtained for all models in both dry and saline environments. The graphical analysis reveals that the UD613 model consistently exhibits the highest stress values across both testing conditions, whereas the V4 and GF3 models maintain significantly lower stress levels, suggesting enhanced mechanical buffering.

In terms of strain, the UD613 geometry reaches its peak value in the dry environment; however, the V5 model stands out with the highest strain levels in the saline environment. When evaluating material stiffness through elastic modulus values, the V5 model manifests the highest stiffness under dry conditions (154.53 Pa), while the V2 model demonstrates the maximum modulus in the saline environment (188.32 Pa). These trends underscore the critical influence of environmental factors and geometric modifications on the overall stability and mechanical response of the haptic designs. All simulation results obtained are consistent with similar studies in the literature (Ammari et al., 2010; Cabeza-Gil et al., 2021; Remón et al., 2018; Schor et al., 2007).

## 4. Discussion

This study comprehensively evaluated the mechanical stability of three commercial and five geometrically modified intraocular lens (IOL) models under both dry and physiologically relevant saline conditions using finite element analysis (FEA) and experimental validation. The findings provide critical insights into how haptic arm geometry influences IOL mechanical behavior, particularly regarding stress distribution and deformation under compressive loads. Furthermore, this research offers a validated framework for optimizing lens designs to enhance stability and performance in clinical applications.

### 4.1. Mechanical Stability and Clinical Implications

The mechanical stability of an intraocular lens (IOL) within the capsular bag is paramount for maintaining optimal optical performance. Mechanical instability—manifested as tilt, decentration, or rotation—can lead to significant visual disturbances, including induced astigmatism, halos, and diminished contrast sensitivity (Remón et al., 2018). Our findings demonstrate that the UD613 model exhibited the highest compression force and stress values in both dry and saline environments, indicating a stiffer overall mechanical response.

While this increased stiffness may facilitate secure fixation within the capsular bag, the elevated stress concentrations observed in this model raise critical concerns regarding long-term structural integrity. Specifically, such high localized stress may increase the susceptibility to material fatigue or permanent haptic deformation over time. These factors must be carefully balanced to ensure that the haptics provide sufficient capsular bag tension without compromising the structural longevity of the lens.

In contrast, the GF3 model and its V4 variation demonstrated more balanced mechanical responses, characterized by lower stress and strain values while maintaining adequate compression force. The V4 geometry, in particular, exhibited the most favorable mechanical profile across all evaluated parameters, showing minimal deformation—0.0361 mm in dry conditions and $9.12 \times 10^{-2}$ mm in saline—alongside consistently low stress concentrations.

From a clinical perspective, this balanced behavior is highly desirable, as it suggests the lens can maintain stable positioning within the capsular bag without generating excessive localized stresses that might compromise long-term structural integrity. Such stability is particularly critical for premium IOL categories, such as multifocal or toric lenses, where even minor deviations in centration or tilt can significantly impair visual outcomes (Cabeza-Gil et al., 2021). Consequently, the V4 design represents a robust optimization that effectively reconciles the need for secure fixation with the requirement for mechanical longevity.

### 4.2. Comparison with Previously Reported IOL FEM Studies

The findings of this study are consistent with and extend previous research on intraocular lens (IOL) mechanical behavior. Remón et al. (2018) demonstrated that both material properties and haptic design significantly influence IOL mechanical stability, noting that C-loop haptics typically offer more favorable stress distribution compared to plate-haptic designs. Our results further corroborate these findings, as the GF3 model—which features a modified C-loop architecture—demonstrated a more balanced mechanical response than the ALSEE and UD613 models.

By refining this geometry in the V4 variation, we achieved even greater optimization in stress dissipation while maintaining the necessary compression force. This suggests that incremental geometric modifications to established haptic designs can yield superior structural performance, effectively bridging the gap between theoretical modeling and clinical requirements for long-term lens stability.

The findings of this study further expand upon the work of Cabeza-Gil et al. (2021), who investigated the influence of haptic geometry in C-loop IOLs and reported that specific geometric modifications can significantly alter the resulting mechanical response. Their research indicated that increasing haptic angulation effectively reduces axial displacement but simultaneously elevates stress concentrations at the haptic-optic junction.

This study extends these observations by systematically varying multiple geometric parameters—beyond simple angulation—and evaluating their synergistic effects under both dry and physiologically relevant saline conditions. Among the variations tested, the V4 geometry achieved the optimal balance between minimized deformation ($9.12 \times 10^{-2}$ mm in saline) and manageable stress levels ($1.12 \times 10^{-3}$ Pa), aligning with the fundamental design principles proposed by Cabeza-Gil et al. (2021) while offering a more robust solution for saline environments. This demonstrates that a multi-parameter optimization approach is essential for developing haptics that remain stable without compromising material longevity under physiological stress.

The findings of this research also intersect with the work of Schor et al. (2007), who modeled accommodating IOLs as spring systems and demonstrated that haptic mechanical properties directly influence the dynamic lens response during accommodation. While the present study focused on quasi-static loading scenarios, the experimentally and numerically derived elastic modulus values provide high-fidelity input parameters for such dynamic models.

The temperature-dependent behavior observed throughout our simulations—where saline conditions consistently yielded higher deformation and stress levels—aligns with the observations of Ammari et al. (2010) regarding the high sensitivity of polymers to environmental stimuli. This underscores the critical importance of evaluating IOL performance under physiologically relevant conditions; relying solely on room-temperature (23°C) dry testing may lead to a significant underestimation of *in vivo* deformations, potentially resulting in inaccurate predictions of long-term haptic stability.

More broadly, our results resonate with recent advances in ocular biomechanics modeling. Mazinani (2024, 2025a, 2025b, 2025c) demonstrated the utility of finite element methods (FEM) for characterizing corneal biomechanics, including shear wave propagation and the effects of intraocular pressure. These studies underscore the necessity of accurate material modeling and precise boundary condition definition. Our validated approach contributes to the growing body of evidence supporting FEM as a robust and powerful tool for the design and optimization of next-generation ophthalmic devices.

### 4.3. Limitations of the Current Study

While this research provides critical insights into haptic optimization, several limitations must be acknowledged when interpreting the results.

- **Material Modeling:** The assumption of linear elastic, isotropic, and homogeneous material behavior was utilized for computational efficiency and comparative analysis under low-strain conditions. However, this approach does not fully capture the inherent viscoelastic behavior characteristic of acrylic polymers. Future investigations should incorporate more sophisticated hyperelastic or viscoelastic formulations to better reflect time-dependent material responses.

- **Boundary Conditions:** The fixation of haptic proximal ends as cantilever supports simplifies the complex mechanical interaction between the haptics and the capsular bag. The capsular bag is a dynamic, viscoelastic structure that undergoes significant postoperative remodeling. Subsequent models that incorporate the capsular bag as a deformable structure would provide a more anatomically accurate representation of the *in vivo* environment.
- **Dynamic and Environmental Factors:** The quasi-static loading conditions employed do not represent the dynamic forces experienced during rapid eye movements or the accommodation process. Consequently, long-term durability and fatigue behavior cannot be fully assessed from the current data set. Furthermore, while saline conditions were simulated by adjusting the elastic modulus (reducing it to 4.65 MPa), this remains a simplified approximation of the synergistic effects of temperature and hydration.
- **Sample Size and Scope:** Although experimental validation involved five repetitions per condition, increasing the sample size and including a broader range of commercial designs would enhance the statistical power and generalizability of the findings. Finally, this study focused exclusively on mechanical stability; future research should integrate direct measurements of optical performance parameters, such as modulation transfer function (MTF) and Strehl ratio, to correlate mechanical deformation with visual quality.

### 4.4. Future Directions

Building on these findings, several critical research trajectories emerge for the next generation of intraocular lens development:

- **Opto-Mechanical Correlation**: Future studies should directly correlate mechanical stability parameters with optical performance metrics—such as Modulation Transfer Function (MTF) and Strehl ratio—by introducing controlled tilt and decentration into ray-tracing optical simulations.
- **Personalized Ocular Modeling**: Integration with high-resolution imaging technologies (e.g., OCT or Scheimpflug imaging) could facilitate the creation of patient-specific models, allowing for personalized IOL selection based on individual capsular bag morphology.
- **Machine Learning Integration**: The extensive dataset generated through these simulations could serve as a training foundation for machine learning algorithms, enabling the rapid screening and optimization of thousands of geometric haptic variations.
- **Dynamic Fatigue Analysis**: Investigating optimized geometries, such as the V4 model, under cyclic and dynamic loading conditions would provide essential insights into long-term fatigue resistance and material durability.
- **Clinical Validation**: Ultimately, prospective clinical studies that correlate specific design characteristics with postoperative stability metrics and patient-reported visual outcomes are necessary to establish the strongest evidence-based guidelines for IOL design optimization.

## 5. Conclusion

This study comprehensively investigated the mechanical stability of three commercial intraocular lens (IOL) models (ALSEE, GF3, UD613) and five geometric variations (V1–V5) derived from the GF3 model under both dry and saline conditions. By utilizing the finite element method (FEM) with rigorous experimental validation, this research characterized the stress, strain, elastic modulus, and axial displacement of these designs within a simulated intraocular environment.

The core findings of this investigation are summarized as follows:

- **Model Performance**: The UD613 geometry exhibited the highest compression force and peak stress values, indicating a stiffer mechanical response that, while secure, led to elevated stress concentrations. Conversely, the GF3 model demonstrated a more balanced mechanical profile, establishing it as an optimal baseline for geometric refinement.
- **Geometric Impact**: The results clearly demonstrate that even subtle modifications in haptic arm geometry exert direct and statistically significant effects on the mechanical stability and stress dissipation of the lens.
- **Environmental Sensitivity:** The saline environment consistently resulted in higher deformation, strain, and stress values across all models compared to dry conditions. This highlights the critical necessity of evaluating IOLs under physiologically relevant conditions (37°C and hydration) to avoid underestimating *in vivo* mechanical behavior.
- **Optimal Design (V4):** Among all analyzed models, the V4 geometry consistently demonstrated the most favorable mechanical characteristics. It maintained minimal deformation (0.0361 mm in dry; $9.12 \times 10^{-2}$ mm in saline), low strain (0.0091 mm/mm in dry; $8.45 \times 10^{-6}$ mm/mm in saline), and low stress (72.18 Pa in dry; $1.12 \times 10^{-3}$ Pa in saline) while providing adequate compression force. This balanced response suggests that the V4 model is the most suitable architecture for minimizing tilt, decentration, and rotation.

The primary contribution of this research is the provision of a validated simulation framework that enables manufacturers to design IOLs with enhanced structural stability. While the study is limited by the assumptions of linear elastic material behavior and quasi-static loading, it provides a robust foundation for future work. Subsequent research will integrate viscoelastic modeling, machine learning-driven design screening, and dynamic fatigue analysis to further bridge the gap between computational modeling and clinical outcomes.

In conclusion, this research demonstrates that the strategic geometric optimization of haptic arm design—specifically the **V4 geometry**—can significantly enhance IOL mechanical stability, potentially improving long-term visual quality and patient outcomes in cataract surgery.

**Funding:** This research received no specific grant from any funding agency in the public, commercial, or not-for-profit sectors.

**Ethical Approval:** Not applicable.